\documentclass[12pt]{article}
\usepackage{amscd,amssymb,amsmath,latexsym,enumerate}
\usepackage[mathscr]{euscript}
\usepackage{mathrsfs}
\usepackage{epsfig}
\usepackage{fancybox}
\usepackage{verbatim}
\usepackage{tikz}
\usepackage{tikz-cd}
\usepackage{todonotes}
\usepackage{multicol}
\usepackage{graphicx}
\usepackage{mathtools}

\usepackage{color}

\usepackage{xcolor}
\definecolor{MyBlue}{cmyk}{1,0.13,0,0.63}
\definecolor{MyGreen}{cmyk}{0.91,0,0.88,0.52}
\newcommand{\mylinkcolor}{MyBlue}
\newcommand{\mycitecolor}{MyGreen}
\newcommand{\myurlcolor}{black}

\usepackage{hyperref}
\hypersetup{%
  bookmarksnumbered=true,bookmarksopen=false,%
  plainpages=false,
  linktocpage=true,%
  colorlinks=true,breaklinks=true,%
  linkcolor=\mylinkcolor,citecolor=\mycitecolor,urlcolor=\myurlcolor,%
  pdfpagelayout=OneColumn,%
  pageanchor=true,%
}

\title{Spectral localizer for line-gapped non-hermitian systems}

\author{Alexander Cerjan$^1$, Lars Koekenbier$^2$, Hermann Schulz-Baldes$^2$
\\
\\
{\small $^1$Center for Integrated Nanotechnologies, Sandia National Laboratories}
\\
{\small Albuquerque, New Mexico 87185, USA}
\\
{\small $^2$Friedrich-Alexander-Universit\"at Erlangen-N\"urnberg}
\\
{\small Department Mathematik, Cauerstr.~11, D-91058 Erlangen, Germany}
}

\date{ }

\newtheorem{theorem}{Theorem}
\newtheorem{proposition}[theorem]{Proposition}
\newtheorem{lemma}[theorem]{Lemma}
\newtheorem{corollary}[theorem]{Corollary}

\newcommand{\CM}{{\mathbb C}}

\newcommand{\ZM}{{\mathbb Z}}

\newcommand{\Dd}{{\cal D}}

\newcommand{\Oo}{{\cal O}}

\newcommand{\Hh}{{\cal H}}

\newcommand{\one}{{\bf 1}}

\newcommand{\Tr}{\mbox{\rm Tr}}

\newcommand{\od}{{\mbox{\tiny\rm od}}}

\newcommand{\SF}{{\rm Sf}}

\newcommand{\Ch}{{\rm Ch}} 
\newcommand{\Ind}{{\rm Ind}} 
 
\newcommand{\Ran}{{\rm Ran}}

\newcommand{\Sig}{{\rm Sig}}

\newcommand{\ImVar}{s}

\begin{document}

\maketitle

\begin{abstract}
Short-ranged and line-gapped non-hermitian Hamiltonians have strong topological invariants given by an index of an associated Fredholm operator. It is shown how these invariants can be accessed via the signature of  a suitable spectral localizer. This numerical technique is implemented in an example with relevance to the design of topological photonic systems, such as topological lasers.




\end{abstract}



\vspace{1cm}

\section{Overview}
\label{sec-Intro}

In a series of recent works, Terry Loring and one of the authors \cite{LS1,LS2} proved that the integer-valued strong topological invariants of solid state systems can be computed as the signature of suitable finite-volume approximations of the so-called spectral localizer. Roughly stated, the localizer is the sum of the Dirac operator with the Hamiltonian as a topological mass term. This provides a very effective numerical tool for the local computation of these invariants. The technique has been extended to weak invariants \cite{SS1}, spin Chern numbers \cite{DS1}, $\ZM_2$-invariants in presence of real symmetries \cite{DS2} as well as to the detection of local topological data in semimetals \cite{SS2} and metals \cite{CL,acoustic_metamaterial}. All of these works suppose that the Hamiltonian is selfadjoint. It is the purpose of this note to show that the spectral localizer can also be used in non-hermitian topological systems with a line-gap. While the spectral localizer was recently used to study a specific class of non-hermitian phenomena that can manifest in anomalous Floquet topological insulators \cite{LF}, this approach still employed a self-adjoint spectral localizer. The literature on non-hermitian systems has grown very rapidly in the last years, as non-hermitian Hamiltonians are relevant for dissipative, bosonic and photonic systems, among others. There are numerous physics reviews available \cite{OPA,KSUS,DL,BBK,AGU} that contain an abundance of further references.

\vspace{.2cm}

Let us directly outline the construction of the non-hermitian spectral localizer and its main properties, focussing on bounded Hamiltonians $H$ on a $d$-dimensional tight-binding Hilbert space $\Hh=\ell^2(\ZM^d,\CM^L)$ with $L$ internal degrees of freedom. The Hamiltonian is supposed to be short-range in the sense that there is an $\alpha>d+2$ and a constant $C$ such that
\begin{equation}
\label{eq-ShortRange}
\|\langle n|H|m\rangle\|\;\leq\;\frac{C}{1+|n-m|^\alpha}
\;,
\qquad
n,m\in\ZM^d
\;,
\;\;\alpha>d+2
\;.
\end{equation}
The second main assumption is that $H$ has a line-gap along the imaginary axis quantified by 
$$
g\;=\;\inf_{\ImVar\in\mathbb{R}}\|(H^{\ImVar})^{-1}\|^{-1}
\;,
$$
where $H^{\ImVar}=H+\imath{\ImVar}\one$. One can readily check that $g>0$ if and only if $H$ has no spectrum on the imaginary axis. If the resolvent set contains a different straight line, one can shift and rotate the Hamiltonian into the above standard form. The line-gap allows one to define a Riesz projection $P=\oint_\gamma\frac{dz}{2\pi\imath}\,(z\one-H)^{-1}$ onto the spectrum with negative imaginary part by using any path $\gamma$ encircling it. Even though $P$ is merely an idempotent and not necessarily selfadjoint, it is possible that $P$ contains topological content in the form of the so-called strong invariant. Let us introduce this invariant as an index of a Fredholm operator. Later on its connections with more widely used strong Chern numbers will be mentioned. The index is introduced using the (dual) Dirac operator 
$$
D
\;=\;
\sum_{j=1}^d\Gamma_j X_j
\;,
$$ 
where $\Gamma_1,\ldots,\Gamma_d$ form an irreducible selfadjoint representation of the Clifford algebra with $d$ generators and $X_1,\ldots,X_d$ are the selfadjoint position operators on $\Hh=\ell^2(\ZM^d,\CM^L)$. The irreducible representation acts on $\CM^{d'}$ with $d'=2^{\lfloor\frac{d}{2}\rfloor}$ so that $D$ acts on $\Hh\otimes\CM^{d'}$. Note that $D$ has compact resolvent. In the case that $d$ is even, there exists a selfadjoint unitary $\Gamma=\Gamma_{d+1}$ anti-commuting with  $\Gamma_1,\ldots,\Gamma_d$. In a suitable representation, $\Gamma$ is diagonal and $D$ off-diagonal:
$$
\Gamma
\;=\;
\begin{pmatrix}
\one & 0 \\ 0 & -\one
\end{pmatrix}
\;,
\qquad
D\;=\; 
\begin{pmatrix}0 & D_{0}^{*}\\
D_{0} & 0
\end{pmatrix}
\;.
$$
The Hamiltonian $H\cong H\otimes \one$ is naturally extended to $\Hh\otimes\CM^{d'}$. In Section~\ref{sec-FredModule} it will be shown that the short-range Hamiltonian leaves the domain of $D$ invariant and that $[D,H]$ extends to a bounded operator. In other words \cite{GVF,DSW}, a short-range Hamiltonian $H$ is differentiable w.r.t. $D$ and the Dirac operator $D$ specifies a Fredholm module for $H$ (or more precisely the algebra of polynomials in $H$) which is even/odd if $d$ is even/odd. Let us focus on even $d$, then the Dirac phase is introduced as the unitary operator $F_0=D_0|D_0|^{-1}$ (strictly speaking $D_0$ has a $d'$-dimensional kernel, but on this subspace $F_0$ can simply be set to the identity). Then a modification of standard arguments discussed in Section~\ref{sec-FredModule} shows that the restriction $PF_0P^*|_{\Ran(P)}$ of $F_0$ to the Hilbert space $\Ran(P)$ is a Fredholm operator. Its index is referred to as the even strong index pairing:
$$
\Ind\big(PF_0P^*|_{\Ran(P)}\big)
\;.
$$
By construction, it is a homotopy invariant. Moreover, if $H$ is periodic or, more generally, a homogeneous system, then an index theorem \cite{PS} shows that the index pairing is equal to the $d$th Chern number $\Ch_d(P)$ which in turn is equal to the Chern number $\Ch_d(Q)$ of the selfadjoint projection $Q$ onto $\Ran(P)$ (for the latter, see \cite{PeS} or use the homotopy spelled out in Section~\ref{sec-Proof}).

\vspace{.2cm}

As already stated above, this paper is about a non-hermitian generalization of the spectral localizer and the focus will be on even dimension $d$. For a tuning parameter $\kappa>0$, the even non-hermitian spectral localizer is introduced by
\begin{equation}
\label{eq-SpecLocDef}
L_{\kappa}(H)\;=\;  
\begin{pmatrix}-H & \kappa D_{0}^{*}\\
\kappa D_{0} & H^{*}
\end{pmatrix}
\;.
\end{equation}
This operator acts on $\Hh\otimes\CM^{d'}$ and is here written in the grading of $\Gamma$. Note that for selfadjoint $H$ this reduces to the even spectral localizer used in \cite{LS2,SS1}. Clearly one has 
\begin{equation}
\label{eq-SpecLocShift}
L_{\kappa}(H^{\ImVar})\;=\;   L_{\kappa}(H)\,-\,\imath \,\ImVar\,\one
\;.
\end{equation}
This indicates that $L_{\kappa}(H)$ may have a line-gap, a fact that can indeed be confirmed for $\kappa$ sufficiently small (see Theorem~\ref{theo-Main} below). Next let us introduce finite-volume approximations, just as in prior works. Let $(\Hh\oplus\Hh)_\rho$ be the range of the finite-dimensional projection $\chi(|D|\leq \rho)$ and let $\pi_{\rho}:\Hh\oplus\Hh\to(\mathcal{H}\oplus\mathcal{H})_{\rho}$ be the associated surjective partial isometry. Note that $\one_{\rho}=\pi_{\rho}\pi_{\rho}^{*}$ is then the identity on $(\mathcal{H}\oplus\mathcal{H})_{\rho}$. For any operator $A$ acting on $\mathcal{H}\oplus\mathcal{H}$ denote its compression to $(\mathcal{H}\oplus\mathcal{H})_{\rho}$ by $A_{\rho}=\pi_{\rho}A\pi_{\rho}^{*}$. The finite-volume non-hermitian spectral localizer is then given by $L_{\kappa}(H)_\rho$ and denoted $L_{\kappa,\rho}(H)=L_{\kappa}(H)_\rho$.

\begin{theorem}
\label{theo-Main}
Suppose that $H$ is short range and set $N=\max\{\|[D,H]\|,\|[|D|,H]\|\}<\infty$ where $H\cong H\otimes \one$ and $|D|=(D^*D)^{\frac{1}{2}}$ is the absolute value of the Dirac operator. If
\begin{equation}
\label{eq-HypBound}
\kappa\;\leq\;c_\kappa\,\frac{g^{3}}{\|H\|\,N}
\qquad\mbox{and}\qquad c_\rho\,\frac{g}{\kappa}
\Big(1+\frac{\|\Im m(H)\|}{g}\Big)
\;\leq\;\rho
\;,
\end{equation}
for $c_\kappa=\frac{1}{12}$ and $c_\rho=6$, then $L_{\kappa,\rho}(H)$ has a quantitative line-gap on the imaginary axis in the sense that, for all $\ImVar\in\mathbb{R}$,
\begin{equation}
\label{eq-GapSize}
L_{\kappa,\rho}(H^{\ImVar})^{*}L_{\kappa,\rho}(H^{\ImVar})
\;\geq\;  \frac{g^2}{4}\, \boldsymbol{1}_{\rho}
\end{equation}
and 
\begin{equation}
\label{eq-SigInd}
\Ind\big(PF_0P^*|_{\Ran(P)}\big)
\;=\;
\frac{1}{2}\;\Sig(L_{\kappa,\rho}(H))
\;,
\end{equation}
where here the signature denotes the difference of the joint algebraic multiplicities of eigenvalues with positive and negative real parts.
\end{theorem}

Let us make a few comments. First of all, compared with earlier works the second bound in \eqref{eq-HypBound} has a supplementary factor $1+\frac{\|\Im m(H)\|}{g}$ which is needed to control the non-hermitian part of the localizer. It is not needed for the proof of the bound \eqref{eq-GapSize} in Section~\ref{sec-LineGap}, but merely for the proof of the constancy of the signature in Section~\ref{sec-SigConst}. Numerical implementation shows that \eqref{eq-HypBound} is far from optimal, and indeed in applications one rather verifies that the line-gap of $L_{\kappa,\rho}(H)$ is open before confidently using its signature. Let us also stress that the supplementary factor does not alter the invariance of the two bounds \eqref{eq-HypBound} under scaling $H\mapsto\lambda H$ which implies $g\mapsto \lambda g$ and $\kappa\mapsto\lambda \kappa$, so that the condition on $\rho$ remains unchanged. As in all prior works the constants $c_\kappa$ and $c_\rho$ in  \eqref{eq-HypBound} are not optimal, but rather a result of the method of proof and the choices made in the proof. Second of all,  it is, in general, {\it not} sufficient to compute the spectrum of the real part $\Re e( L_{\kappa,\rho}(H))=\frac{1}{2}(L_{\kappa,\rho}(H)+L_{\kappa,\rho}(H)^*)$ because $H$ may be non-normal. However, as in applications one typically only needs to consider relatively small $\rho$ and thus relatively small non-hermitian matrices $L_{\kappa,\rho}(H)$, this is not really a limitation, as show the examples in Section~\ref{sec-Numerics}. Third of all, let us mention that Appendix~\ref{sec-Signature} describes two efficient techniques to access the signature, one via spectral flow and one by a Routh-Hurewitz theorem. Finally, let us note that in the earlier works \cite{LS2,DSW} only the constant $\|[D,H]\|$ entered in the bounds, while here also the norm of the commutator $[|D|,H]$ is of relevance. Its boundedness can also be shown if $H$ satisfies the short-range condition \eqref{eq-ShortRange}, see Section~\ref{sec-FredModule}. The Fredholm module is then referred to as Lipshitz regular. An alternative way to guarantee the Lipshitz regularity automatically is to replace the Dirac operator $D$ by $D(\one+D^2)^{-\beta}$ for some $\beta>0$  \cite{Kaad2020,SS3}. The index pairing remains unchanged during the homotopy $\beta'\in[0,\beta]\mapsto D(\one+D^2)^{-\beta'}$. Clearly, also the signature in \eqref{eq-SigInd} does not change as long as $\beta$ is sufficiently small. 

\vspace{.2cm}

Up to now, only the case of even dimension $d$ was considered. For odd $d$ and hermitian systems, a strong topological invariant is only defined if $H$ has a chiral symmetry of the form $JHJ=-H$ where $J=J^*=J^{-1}$. Then there are odd index pairings and odd Chern numbers \cite{PS} which can be computed with an odd spectral localizer \cite{LS1}. In Section~\ref{sec-OddDim} it will be explained that this story directly transposes to the study of non-hermitian line-gapped chiral Hamiltonians.

\section{Numerical implementation}
\label{sec-Numerics}

\begin{figure*}[h!]
  \includegraphics[width=\linewidth]{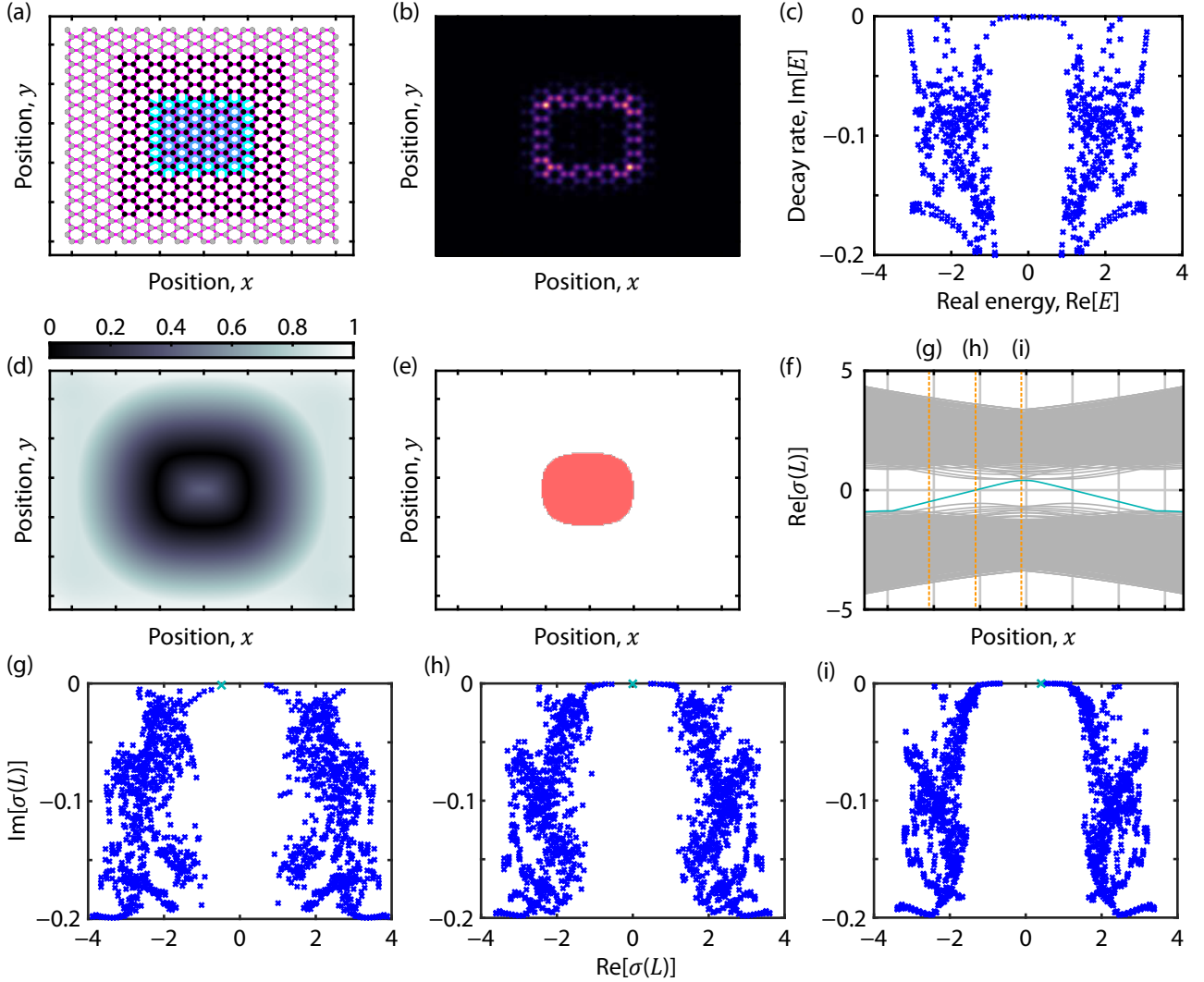}
  \caption{(a) Diagram of the tight-binding heterostructure consisting of a topological insulating lattice in the center surrounded by a trivial insulator whose perimeter contains loss. For the topological insulator, $M = 0$, $t_c = 0.5$, and $\phi = \pm\frac{\pi}{2}$. For the trivial insulator, $M = 0.5\sqrt{3}$ and $t_c = 0$. Both lattices have $t=1$. The black vertices are lossless, while the gray vertices have $\mu = 0.2$. (b) Local density of states for this heterostructure at $E = 0$. (c) Full complex spectrum of the heterostructure. (d) The localizer gap given by the smallest of the absolute values of the real parts of the eigenvalues of $L_{\kappa}(H,x,y)$, namely $\textrm{min}\,|\Re e(\sigma(L_{\kappa}(H,x,y)))|$. (e)~Spatially resolved local index. The red region shows where the index is non-trivial and equal to $1$. (f)~Real part of the spectral flow of $L_{\kappa,\rho}(H,x,0)$ as a function of position in $x$. The eigenvalue responsible for the index change is highlighted in teal. (g),(h),(i) Full complex spectrum of $L_{\kappa}(H,x,0)$ for three different choices of $x$; the choices of $x$ are indicated by orange dashed lines in (f). Again, the eigenvalue responsible for the change in index is shown in teal. The scales of (a),(b),(d),(e) and (f) are the same, and $\kappa = 0.1$ for all spectral localizer calculations.} 
  \label{fig:1}
  \end{figure*}

To provide an explicit example of the utility of the non-hermitian generalization of the spectral localizer, let us consider a finite heterostructure comprised of two lattices in different topological phases. More specifically, suppose given a Haldane model over a bi-partite honeycomb lattice $\Gamma=\Gamma_A\cap\Gamma_B$ \cite{Hal}, whose tight-binding model is 
\begin{align}
H  
\;=\; 
&
\sum_{n_A,n_B} 
 \big(M\,|n_A\rangle\langle n_A|\,-\,M\,|n_B\rangle\langle n_B|\big) 
 \,-\, 
 t \sum_{\langle n_A,m_B\rangle} 
 \big(|n_A\rangle\langle m_B|\,+\,|m_B\rangle\langle n_A|\big) 
\nonumber
\\
& \,-\,
 t_c\sum_{\alpha=A,B} \sum_{\langle\!\langle n_\alpha,m_\alpha\rangle\!\rangle}  
 \big(e^{\imath\phi(n_\alpha,m_\alpha)}\, |n_\alpha\rangle\langle m_\alpha|\,+\,e^{-\imath\phi(n_\alpha,m_\alpha)}\,|m_\alpha\rangle\langle n_\alpha|\big) 
 \label{eq-haldaneH}
\end{align}
Here the first sum runs over all sites in the lattice and is a staggered potential giving the $A$ and $B$ lattices opposite on-site energies $M$ and $-M$, the second sum is a kinetic energy with nearest neighbor coupling coefficient $t$ and the third sum 
is over next-nearest-neighbor pairs and has a direction-dependent phase factor that breaks time-reversal symmetry with a periodic magnetic field, namely $\phi(n_\alpha,m_\alpha)=\pm\phi$ with a geometrically chosen sign \cite{Hal}. The Hamiltonian is known to have a spectral gap at $0$ with a topological Fermi projection $P$  for $M\ll t_c$, and it is a topologically trivial insulator for $t_c\ll M$ (see \cite{Hal} for the phase diagram).  Furthermore, the model can be made lossy with absorption strength $\mu$ if $M$ is replaced by $M \mp \imath\mu$ on the $A$ and $B$ sublattices respectively. Altogether, the heterostructure is made up of a topological Haldane model in the central part, surrounded first by a ring of trivial insulator and then a ring of a lossy trivial insulator, see Fig.~\ref{fig:1}(a). 

\vspace{.2cm}

The choice of loss distribution around the lattice's perimeter is guided by analogy to photonic systems, as such systems are  one of the most common platforms where non-hermiticity can manifest in topological materials characterized by line-gaps \cite{OPA,BBK,cerjan_nanophot}. Unlike electronic systems, for which free space is a trivial insulator, many photonic systems will radiate into their surrounding free-space environment. This radiation can be considered by surrounding a region of interest using an absorbing boundary condition, such as perfectly matched layers \cite{taflove}, which necessarily makes the full system non-hermitian. Heuristically, the purpose of the absorbing boundary condition is to replicate the infinite extent of the environment in a finite simulation domain without introducing spurious reflections.

\vspace{.2cm}

The local density of states (LDOS) of the heterostructure at energy $E=0$ is shown Fig.~\ref{fig:1}(b) and the complete spectrum in Fig.~\ref{fig:1}(c), for parameter values as described there. Note that essentially the only eigenvalues with very small imaginary part are the surface states in the topological central part, as they are separated from the lossy region by the trivial insulator which has an energy gap at $E=0$. 

\vspace{.2cm}

The different local topologies can be identified in the finite non-hermitian heterostructure using the local topological invariant (local marker) given in \eqref{eq-SigInd} with a position shift $x,y$ of the Dirac operator, namely by the half-signature of
$$
L_\kappa(H,x,y)
\;=\;
\begin{pmatrix}-H & \kappa \big((X-x)-\imath (Y-y)\big)\\
\kappa \big((X-x)+\imath (Y-y)\big) & H^{*}
\end{pmatrix}
\;,
$$
where $X$ and $Y$ are the two position operators (denoted by $X_1$ and $X_2$ above) and there is no finite size restriction as all matrices are finite here. The size of the line-gap of $L_\kappa(H,x,y)$ at $\Re e(E)=0$ is shown in Fig.~\ref{fig:1}(d) and the value of the half-signature as defined in Theorem~\ref{theo-Main} in Fig.~\ref{fig:1}(e). This is computed by the spectral flow method described in Appendix~\ref{sec-Signature} by using the path $t\in[0,T]\mapsto L_\kappa(H,x+t,y)$ and the fact that $\Sig( L_\kappa(H,x+T,y))=0$ for sufficiently large $T$, say so that $x+T$ lies outside of the boundary of the heterostructure. An example of a spectral flow diagram for the real part of the spectrum is given in Fig.~\ref{fig:1}(f) where the eigenvalue responsible for the signature change is readily visible. To complement the picture, Fig.~\ref{fig:1}(g), (h)  and (i) show the full complex spectrum of $L_\kappa(H,x,y)$ for three different values of $x$. Here Fig.~\ref{fig:1}(g) and (i) correspond to the exterior and central regions where one clearly sees the line-gap at $\Re e(E)=0$ which corresponds to part of the statement of Theorem~\ref{theo-Main} for the topological and trivial insulator respectively. Finally let us note that, as expected, in Fig.~\ref{fig:1}(d) the local invariant changes near the interface between the two lattices due to the presence of the chiral interface-localized states visible in Fig.~\ref{fig:1}(b).

\vspace{1cm}

\section{Fredholm properties}
\label{sec-FredModule}

\begin{lemma}
\label{lem-FredMod}
If $H$ satisfies the short-range condition \eqref{eq-ShortRange}, then $H$ leaves the domain of $D$ invariant and the commutators $[D,H]$ and $[|D|,H]$ extend to bounded operators.
\end{lemma}

\noindent {\bf Proof.}
As $D^2=\sum_{j=1}^dX_j^2=X^2$, its domain is $\Dd(D)=\{\psi\in\Hh\otimes\CM^{d'}\,:\,\sum_{n\in\ZM^d}|n|^2\|\psi_n\|^2<\infty\}$. Now
\begin{align*}
\sum_{n\in\ZM^d}|n|^2\|(H\psi)_n\|^2
&
\;=\;
\sum_{n,m,k\in\ZM^d}\psi_k^*\langle k|H^*|n\rangle\,|n|^2\,\langle n|H|m\rangle \,\psi_m
\\
&
\;\leq\;
\sum_{n,m,k\in\ZM^d}\|\psi_k\|\,\frac{C}{1+|n-k|^\alpha}\,|n|^2\,\frac{C}{1+|n-m|^\alpha}\,\|\psi_m\|
\\
&
\;\leq\;
\sum_{n,m,k\in\ZM^d}\|\psi_k\|^2\,\frac{C}{1+|n-k|^\alpha}\,|n|^2\,\frac{C}{1+|n-m|^\alpha}
\\
&
\;\leq\;
\sum_{k\in\ZM^d}|k|^2 \|\psi_k\|^2\,\sup_{k'\in\ZM^d}\frac{1}{1+|k'|^2}
\sum_{n,m\in\ZM^d}
\,\frac{C}{1+|n-k'|^\alpha}\,|n|^2\,\frac{C}{1+|n-m|^\alpha}
\\
&
\;\leq\;
\Big(\sum_{k\in\ZM^d}|k|^2 \|\psi_k\|^2\Big)\sup_{k'\in\ZM^d}\frac{1}{1+|k'|^2}
\sum_{n\in\ZM^d}
\,\frac{C'(|n-k'|^2+|k'|^2)}{1+|n-k'|^\alpha}
\;,
\end{align*}
which is bounded for $\psi\in\Dd(D)$ as $\alpha-2>d$. Hence $H\psi\in\Dd(D)$. Next note that
$$
\langle n|[D,H]|m\rangle
\;=\;
D(n)\,\langle n|H|m\rangle
\,-\,\langle n|H|m\rangle\,D(m)
\;=\;
D(n-m)\,\langle n|H|m\rangle
\;,
$$
where $D(n)=\sum_{j=1}^dn_j\Gamma_j=\langle n|D|n\rangle$. One has $\|D(n)\|\leq\sqrt{d}\,|n-m|$ by the Cauchy-Schwarz inequality. Furthermore, it was used that $H\cong H\otimes\one$ commutes with the $\Gamma_j$'s. Estimating the norm using Holmgren's bound (which contains the maximum of two expressions, but they are bounded in the same manner) gives
\begin{align*}
\|[D,H]\|
&
\;\leq\;
\sup_{n\in\ZM^d}\sum_{m\in\ZM^d}
\|D(n-m)\|\,\|\langle n|H|m\rangle\|
\;\leq\;
\sup_{n\in\ZM^d}\sum_{m\in\ZM^d}
\sqrt{d}\,|n-m|\,
\frac{C}{1+|n-m|^\alpha}
\;,
\end{align*}
which is bounded because $\alpha>d+1$. In order to bound the second commutator, let us set $F=D|D|^{-1}$ and use
$$
[|D|,H]
\;=\;
[F^*D,H]
\;=\;
[F^*,H]D
\,+\,
F^*[D,H]
\;.
$$
As $F$ is unitary, it is hence sufficient to show that $[F^*,H]D$ extends to a bounded operator. Let us write out the matrix elements using $F(n)=\langle n|F|n\rangle$:
$$
\langle n|[F^*,H]D|m\rangle
\;=\;
(F(n)^*-F(m)^*)D(m)\,\langle n|H|m\rangle
\;.
$$
Next let us note the bound
$$
\|F(n)-F(m)\|
\;\leq\;
\sqrt{d}\,\big|\tfrac{n}{|n|}-\tfrac{m}{|m|}\big|
\;\leq\;
2\,\sqrt{d}\;
|n-m|\;\min\{
\tfrac{1}{|n|},\tfrac{1}{|m|}\}
\;,
$$
which can be checked using the Cauchy-Schwarz inequality as above. Hence again appealing to Holmgren's bound gives
\begin{align*}
\|[F^*,H]D\|
&
\;\leq\;
\sup_{n\in\ZM^d}\sum_{m\in\ZM^d}
\|F(n)-F(m))\|\,\|D(m)\|\,\|\langle n|H|m\rangle\|
\\
&
\;\leq\;
\sup_{n\in\ZM^d}\sum_{m\in\ZM^d}
2\,d\;
|n-m|\;\min\{
\tfrac{1}{|n|},\tfrac{1}{|m|}\}
\,|m|\,
\frac{C}{1+|n-m|^\alpha}
\;,
\end{align*}
which is bounded as can be seen by splitting the sum in $|m|<|n|$ and $|m|\geq |n|$. 
\hfill $\Box$

\begin{corollary}
\label{coro-FredProp}
If $H$ satisfies the short-range condition \eqref{eq-ShortRange}, then the commutators $[P,F_0]$ and $[P^*,F_0]$ are compact.
\end{corollary}

\noindent {\bf Proof.}
Given the results of Lemma~\ref{lem-FredMod}, the compactness of $[P,F_0]$ follows directly from the standard arguments ({\it e.g.} in Theorem~10.1.4 in \cite{DSW} which at no point pends on the selfadjointness of $H$; note that the $F$ there is denoted by $F_0$ here).
\hfill $\Box$

\vspace{.2cm}

Now let us construct Fredholm operators from $P$ and $F_0$. For this purpose, let us set 
$$
R
\;=\;
P(\one-(P^*-P)^2)^{-\frac{1}{2}}
\;,
$$
which exists as $-(P^*-P)^2=|P^*-P|^2\geq 0$. One furthermore readily checks that $[P,(P^*-P)^2]=0$ so that $(P-P^*)^2$ and functions thereof leave $\Ran(P)$ and $\Ran(P^*)$ invariant. Then the (orthogonal) projection $Q$ onto the range of $P$ is given by
$$
Q
\;=\;
RR^*
\;=\;P(\one+(P^{*}-P)(P-P^{*}))^{-1}P^*
\;=\; P(P^*P)^{-1}P^*
\;.
$$
%

\begin{proposition}
\label{prop-FredProp2}
If $H$ satisfies the short-range condition \eqref{eq-ShortRange}, then $RF_0R^*+(\one-RR^*)$ and $PF_0P^*|_{\Ran(P)}$ are Fredholm operators and their indices are equal.
\end{proposition}

\noindent {\bf Proof.} First let us note that
$$
[R,F_0]
\;=\;
[P,F_0](\one-(P^{*}-P)^2)^{-\frac{1}{2}}
\,+\,
P[(\one-(P^{*}-P)^2)^{-\frac{1}{2}},F_0]
$$
The first summand is compact by Corollary~\ref{coro-FredProp}. To verify the compactness of the second summand, one can use the norm convergent Riemann integral
$$
(\one-(P^{*}-P)^2)^{-\frac{1}{2}}
\;=\;
\int^\infty_0\frac{d\lambda}{\lambda^{\frac{1}{2}}}\,
(\lambda+\one-(P^{*}-P)^2)^{-1}
\;,
$$
which shows that
$$
[(\one-(P^{*}-P)^2)^{-\frac{1}{2}},F_0]
\;=\;
\int^\infty_0\frac{d\lambda}{\lambda^{\frac{1}{2}}}\,
(\lambda+\one-(P^{*}-P)^2)^{-1}
[(P^{*}-P)^2,F_0]
(\lambda+\one-(P^{*}-P)^2)^{-1}
$$
is also compact, again by Corollary~\ref{coro-FredProp}. Now let us set $T=RF_0R^*+(\one-RR^*)$. Then
\begin{align*}
T^*T
&
\;=\;
RF_0^*R^*RF_0R^*+(\one-RR^*)
\\
&
\;=\;
RR^*F_0^*RF_0R^*+R[F_0^*,R^*]RF_0R^* +(\one-RR^*)
\\
&
\;=\;
RR^*F_0^*F_0RR^*+RR^*F_0^*[R,F_0]R^*+ R[R,F_0]^*RF_0R^* +(\one-RR^*)
\\
&
\;=\;
\one+QF_0^*[R,F_0]R^*+ R[R,F_0]^*RF_0R^*
\;.
\end{align*}
As the last two summands are compact, this implies the desired Fredholm property of $T$. As the two summands in $T$ are orthogonal and one is trivial (given by $\one-RR^*=\one-Q$ with vanishing index), one concludes that also $RF_0R^*|_{\Ran(Q)}$ is Fredholm with same index as $T$. Furthermore, 
\begin{align*}
\Ind\big(RF_0R^*+(\one-RR^*)\big)
&
\;=\;
\Ind\big(RF_0R^*|_{\Ran(Q)}\big)
\\
&
\;=\;
\Ind\big(P(\one-(P^*-P)^2)^{-\frac{1}{2}}
F_0(\one-(P^*-P)^2)^{-\frac{1}{2}}P^*|_{\Ran(P)}\big)
\\
&
\;=\;
\Ind\big((\one-(P^*-P)^2)^{-\frac{1}{2}}P
F_0P^*(\one-(P^*-P)^2)^{-\frac{1}{2}}|_{\Ran(P)}\big)
\\
&
\;=\;
\Ind\big(PF_0P^*|_{\Ran(P)}\big)
\;,
\end{align*}
because $(\one-(P^*-P)^2)^{-\frac{1}{2}}$ is invertible and leaves $\Ran(P)$ and $\Ran(P^*)$ invariant. This proves the claim. 
\hfill $\Box$

\section{Line-gap of the spectral localizer}
\label{sec-LineGap}

This section is entirely devoted to the proof of \eqref{eq-GapSize} under the condition that \eqref{eq-HypBound} holds. While the strategy is similar to earlier arguments \cite{LS2,DSW}, there are some novel difficulties linked to the non-hermitian nature of the Hamiltonian and the spectral localizer that we hope to address clearly in this section. For this reason we merely restrict to the proof of \eqref{eq-GapSize}, even though the very same strategry will be expanded (and thus to some extend repeated) to a proof of the constancy of $\Sig(L_{\kappa,\rho}(H))$ in the next Section~\ref{sec-SigConst}. Let us start from
\begin{align*}
L_{\kappa,\rho}(H^{\ImVar})^{*}L_{\kappa,\rho}(H^{\ImVar})\;=\; 
& (L_{\kappa,\rho}(H)-\ImVar \imath \boldsymbol{1}_{\rho})^{*}(L_{\kappa,\rho}(H)-\ImVar \imath \boldsymbol{1}_{\rho})
\\
\;=\; &  L_{\kappa,\rho}(H)^{*}L_{\kappa,\rho}(H)\,+\,\ImVar^{2}\boldsymbol{1}_{\rho}\,-\,
2\ImVar\,\Im m(L_{\kappa,\rho}(H))\\
\;=\; &  L_{\kappa,\rho}(H)^{*}L_{\kappa,\rho}(H)\,+\,\ImVar^{2}\boldsymbol{1}_{\rho}\,-\,2\ImVar\,(\Im m(-H_\rho)\oplus\Im m(H^{*}_\rho))
\;,
\end{align*}
where $\Im m(A)=\frac{1}{2\imath}(A-A^*)$ is the imaginary part of the operator $A$. Hence one has for all $\ImVar$
\begin{align}
L_{\kappa,\rho}(H^{\ImVar})^{*}L_{\kappa,\rho}(H^{\ImVar})
\;\geq \;  
L_{\kappa,\rho}(H)^{*}L_{\kappa,\rho}(H)+(\ImVar^{2}\,-\,2\,\|\Im m(H)\|\cdot|\ImVar|)\boldsymbol{1}_{\rho}
\;.
\label{eq-Sdrop}
\end{align}
Note that $\ImVar^{2}-2\,\|\Im m(H)\|\cdot|\ImVar|\geq0$ for all $\ImVar$
with $|\ImVar|\geq2\,\|\Im m(H)\|$. Thus for the proof of \eqref{eq-GapSize} it is sufficient to show that, for all $|\ImVar|\leq 2\,\|\Im m(H)\|$,
\begin{align}
L_{\kappa,\rho}(H^{\ImVar})^{*}L_{\kappa,\rho}(H^{\ImVar})
\;\geq \; 
\tfrac{g^2}{4}\,\boldsymbol{1}_{\rho}
\;.
\label{eq-Lbound}
\end{align}
Multiplying out, one finds
\begin{align*}
L_{\kappa,\rho} & (H^{\ImVar})^{*}  L_{\kappa,\rho}(H^{\ImVar})
\\
&
=\;  \kappa^{2}D_\rho^{2}\;+\;|(-H_\rho^{\ImVar})\oplus (H_\rho^{\ImVar})^*|^2
\; +\;\kappa\big(D_\rho((-H_\rho^{\ImVar})\oplus (H^{\ImVar}_\rho)^*)+((-H_\rho^{\ImVar})^*\oplus H_\rho^{\ImVar})D_\rho\big)
\\
&
=\; 
 \kappa^{2}\pi_{\rho}D^{2}\pi_{\rho}^{*}\;+\;\pi_{\rho} ((-H^{\ImVar})^*\oplus H^{\ImVar})\one_\rho ((-H^{\ImVar})\oplus (H^{\ImVar})^*)\pi_{\rho'}^{*}
\;+\;
\kappa\pi_{\rho}\begin{pmatrix}0 & [H,D_{0}]^{*}\\{}
[H,D_{0}] & 0
\end{pmatrix}\pi_{\rho}^{*}
\;,
\end{align*}
where the last step is based on the algebraic identity
\begin{align*}
D((-H^{\ImVar})\oplus (H^{\ImVar})^*)+((-H^{\ImVar})^*\oplus H^{\ImVar})D
\;=\; 
\begin{pmatrix}0 & [H,D_{0}]^{*}\\{}
[H,D_{0}] & 0
\end{pmatrix}
\;.
\end{align*}
The first two summands in $|L_{\kappa,\rho} (H^{\ImVar})|^2$ are non-negative and on each a quantitative (positive) lower bound will be proved below such that the sum of the two is strictly positive; the third summand will then be shown to be a perturbation that does not spoil the positivity. For that purpose, let us use an even differentiable function $G_{\rho}\colon\mathbb{R}\to[0,1]$ constructed in references \cite{LS2,DSW} which satisfies  $G_{\rho}(x)=1$ for all $|x|\leq\frac{1}{2}\rho$ and $G_{\rho}(x)=0$ for all $|x|\geq\rho$, and for which, moreover, the Fourier transform $\widehat{G'_{\rho}}\colon\mathbb{R}\to\mathbb{R}$ of the derivative $G'_{\rho}$ has an $L^{1}$-norm bounded by $8\rho^{-1}$. Then (by Lemma 10.15 in \cite{GVF}) one has for all self-adjoint operators $A$ and bounded operators $B$
\begin{align}
\|[G_{\rho}(A),B]\|
\;\leq \; \tfrac{8}{\rho}\;\|[A,B]\|
\;.
\label{eq-TuningBound}
\end{align}
With this function, one can bound the first summand by showing
\begin{align*}
\kappa^{2}\pi_{\rho}D^{2}\pi_{\rho}^{*}\;\geq \;  g^{2}\pi_{\rho}(\one-G_{\rho}(D)^{2})\pi_{\rho}^{*}
\;.
\end{align*}
Indeed, using a rough version of the second hypothesis in \eqref{eq-HypBound}, one has $\kappa^{2}\geq g^{2}(\frac{1}{2}\rho)^{-2}$ so that the function $G_{\rho}$ satisfies for $x\in[\frac{1}{2}\rho,\rho]$:
\begin{align*}
\kappa^{2}x^{2}
\;\geq \;  
g^{2}(\tfrac{1}{2}\rho)^{-2}x^{2}
\;\geq\; 
g^{2}
\;\geq\; 
g^{2}(\one-G_{\rho}(x)^{2})
\;,
\end{align*}
since $0\leq G_{\rho}(x)\leq1$. On the other hand, for $x\in [0,\frac{1}{2}\rho]$ the bound holds trivially since there $\one-G_{\rho}(x)^{2}=0$. In the second summand, one uses the lower bound $\one_\rho\geq G_\rho(D)^2$ implying
\begin{align*}
|(-H_\rho^{\ImVar})\oplus (H_\rho^{\ImVar})^*|^2
\;\geq\;
&
\pi_{\rho} ((-H^{\ImVar})^*\oplus H^{\ImVar}) G_\rho(D)^2 ((-H^{\ImVar})\oplus (H^{\ImVar})^*){\pi_{\rho}^{*}}
\\
\;=\;&
\pi_{\rho}G_\rho(D) |(-H^{\ImVar})^*\oplus H^{\ImVar}|^2 G_\rho(D) {\pi_{\rho}^{*}}
\\
&
\,+\,
\pi_{\rho} ((-H^{\ImVar})^*\oplus H^{\ImVar}) G_\rho(D) [G_\rho(D),((-H^{\ImVar})\oplus (H^{\ImVar})^*)]{\pi_{\rho}^{*}}
\\
&
\,+\,
\pi_{\rho} [((-H^{\ImVar})^*\oplus H^{\ImVar}), G_\rho(D)] ((-H^{\ImVar})\oplus (H^{\ImVar})^*)G_\rho(D){\pi_{\rho}^{*}}
\end{align*}
Here the first summand can be bounded below by $|(-H^{\ImVar})^*\oplus H^{\ImVar}|^2\geq g^2\,\one$, using the line-gap. Collecting these above lower bounds shows
$$
L_{\kappa,\rho} (H^{\ImVar})^{*}  L_{\kappa,\rho}(H^{\ImVar})
\;\geq\;  
g^{2}\one_{\rho}
\,+\,
E\;,
$$
with an error term given by
\begin{align*}
E
\;=\;
&
\kappa\pi_{\rho}\begin{pmatrix}0 & [H,D_{0}]^{*}\\{}
[H,D_{0}] & 0
\end{pmatrix}\pi_{\rho}^{*}
\\
&
\;+\;
\pi_{\rho} ((-H^{\ImVar})^*\oplus H^{\ImVar}) G_\rho(D) [G_\rho(D),((-H^{\ImVar})\oplus (H^{\ImVar})^*)]\pi_{\rho}^{*}
\\
&
\,+\,
\pi_{\rho} [((-H^{\ImVar})^*\oplus H^{\ImVar}), G_\rho(D)] ((-H^{\ImVar})\oplus (H^{\ImVar})^*)G_\rho(D)\pi_{\rho}^{*}
\;.
\end{align*}
Note that $G_\rho$ is an even function and $|D|^{2}=D^{2}$, so that one can replace $G_{\rho}(D)= G_{\rho}(|D|)$ which is diagonal in the $2\times 2$ grading.  Hence
\begin{align*}
[G_\rho(D),((-H^{\ImVar})\oplus (H^{\ImVar})^*)]
&
\;=\;
[G_\rho(|D_0|),(-H^{\ImVar})]\oplus[G_\rho(|D_0^*|), (H^{\ImVar})^*)]
\\
&
\;=\;
[H,G_\rho(|D_0|)]\oplus[H,G_\rho(|D_0^*|)]^*
\;.
\end{align*}
(Note that for the particular choice of $D$ made here one actually has $|D_0^*|=|D_0|$.) Therefore using $\|G_\rho(D)\|\leq 1$ and then \eqref{eq-TuningBound}, one has
\begin{align*}
\|E\|
&
\;\leq\;\kappa\,\|[H,D_0]\|\;+\;
2\,\|H^{\ImVar}\|\,\max\big\{\|[H,G_{\rho}(|D_0|)]\|,\|H,G_{\rho}(|D_0^*|)]\|\big\}
\\
&
\;\leq\;\kappa\,\|[H,D_0]\|\;+\;
\tfrac{16}{\rho}\,\|H^{\ImVar}\|\,\max\big\{\|[H,|D_0|]\|,\|H,|D_0^*|]\|\big\}
\;.
\end{align*}
Finally let us  use $\|H^{\ImVar}\|\leq\|H\|+|s|\leq \|H\|+2\,\|\Im m(H)\|\leq3\,\|H\|$ (note that the factor $3$ can be omitted if $H$ is selfadjoint, improving the bound below). Then using the quantity $N$ introduced in statement of Theorem~\ref{theo-Main} and the bound $\frac{1}{\rho}\leq \frac{\kappa}{c_\rho g}$ following from \eqref{eq-HypBound}, one deduces
\begin{align}
\|E\|
&
\,\leq\,
\big(\kappa+3\,\|H\|\,\tfrac{16}{\rho}\big)N
\,\leq\,
\kappa\big(1+\tfrac{\|H\|}{g}\,{\tfrac{48}{c_\rho}}\big)N
\,\leq\,
\kappa N \tfrac{\|H\|}{g} {\big(1+\tfrac{48}{c_\rho}\big)}
\,\leq\,
c_\kappa {\big(1+\tfrac{48}{c_\rho}\big)}g^2
\;,
\label{eq-EBound}
\end{align}
due to $\|H\|\geq g$ and \eqref{eq-HypBound}. Now $c_\kappa{(1+\tfrac{48}{c_\rho})}=\frac{3}{4}$, so combining with the above, one deduces \eqref{eq-Lbound} for all $|\ImVar|\leq 2\,\|\Im m(H)\|$.

\section{Constancy of the signature}
\label{sec-SigConst}

It is the object of this section to prove that the signature $\Sig(L_{\kappa,\rho}(H))$ does not change with $\kappa$ and $\rho$, as long as the bounds \eqref{eq-HypBound} hold. For the changes in $\kappa$, this follows directly from the results of Section~\ref{sec-LineGap}, on the other hand changing $\rho$ means changing the size of the matrix which is not a continuous procedure. To address the issue, it will be shown as in \cite{LS2,DSW} that the Hamiltonian can be tampered down away from the origin without changing the signature. Once the corresponding path of tampered spectral localizers is constructed, it is then again sufficient to shown that the line-gap remains open along the path because then there is no spectral flow across the imaginary axis so that the signature remains constant.  This will be achieved by a suitable modification of the arguments of Section~\ref{sec-LineGap}. In particular, the objects and stated bounds of the last section will be freely used. Let us begin by introducing the family of functions  $G_{\rho,\lambda}(x)= (1-\lambda)+\lambda G_{\rho}(x)$ for all {$\lambda \in [0,1]$} and then set
\begin{align*}
L_{\kappa,\rho,\rho'}(H;\lambda)
\;=\; 
& 
\kappa\pi_{\rho'}D\pi_{\rho'}^{*}
\;+\;
\pi_{\rho'}G_{\rho,\lambda}(D)((-H)\oplus (H)^*)G_{\rho,\lambda}(D)\pi_{\rho'}^{*}
\;,
\end{align*}
which is an operator acting on $(\mathcal{H}\oplus\mathcal{H})_{\rho'}$. This formula clearly shows that the Hamiltonian is redressed. One has $L_{\kappa,\rho,\rho'}(H;0)=  L_{\kappa,\rho'}(H)$ and $L_{\kappa,\rho,\rho'}(H,1)=\kappa\pi_{\rho',\rho}D\pi_{\rho',\rho}^{*}+L_{\kappa,\rho,\rho}(H,1)$, where $\pi_{\rho',\rho}$ is the partial isometry onto the subspace of $\Ran(\chi(|D|\leq\rho'))$ that is orthogonal to $\Ran(\chi(|D|\leq\rho))$.  One finds by essentially the same argument leading to \eqref{eq-Sdrop} that
\begin{align*}
\big(L_{\kappa,\rho,\rho'}&(H;\lambda)-\imath\ImVar \one_{\rho'}\big)^*\big(L_{\kappa,\rho,\rho'}(H;\lambda)-\imath\ImVar\one_{\rho'}\big)
\\
&
\,\geq \,  
L_{\kappa,\rho,\rho'}(H;\lambda)^*L_{\kappa,\rho,\rho'}(H;\lambda)
+
(\ImVar^{2}\,-\,2\,\|\Im m (H)||\cdot|\ImVar|)\one_{\rho'}
\,.
\end{align*}
Again this shows that it is sufficient to deal with $|\ImVar|\leq 2\|\Im m(H)\|$. For such $\ImVar$, let us compute again in a similar manner as in Section~\ref{sec-LineGap}, but with a few more algebraic manipulations,
\begin{align}
\big(L_{\kappa,\rho,\rho'}&(H;\lambda)-\imath\ImVar\one_{\rho'}\big)^*\big(L_{\kappa,\rho,\rho'}(H;\lambda)-\imath\ImVar\one_{\rho'}\big)
\nonumber
\\
\;=\; 
& \kappa^{2}\pi_{\rho'}D^{2}\pi_{\rho'}^{*}\,+\,s^{2}\pi_{\rho'}(\one-G_{\rho,\lambda}(D)^{4})\pi_{\rho'}^{*}
\label{eq-Contri1}
\\
&
\;+\;\pi_{\rho'}G_{\rho,\lambda}(D)((-H^{\ImVar})^*\oplus H^{\ImVar})G_{\rho,\lambda}^{2}(D)((-H^{\ImVar})\oplus (H^{\ImVar})^*)G_{\rho,\lambda}(D)\pi_{\rho'}^{*}
\label{eq-Contri2}
\\
& 
\,+\,\kappa\,\pi_{\rho'}G_{\rho,\lambda}(D)\begin{pmatrix}0 & [H,D_{0}]^{*}\\{}
[H,D_{0}] & 0
\end{pmatrix}G_{\rho,\lambda}(D)\pi_{\rho'}^{*}
\label{eq-Contri3}
\\
&
+\;2s\pi_{\rho'}G_{\rho,\lambda}(D)\,\Im m\big[((-H)^{*}\oplus H)(\one-G_{\rho,\lambda}(D)^{2})\big]G_{\rho,\lambda}(D)\pi_{\rho'}^{*}\;.
\label{eq-Contri4}
\end{align}
It is here important that in \eqref{eq-Contri2} appears $H^\ImVar$ and not just $H$, because in this manner the line gap of $H$ can be used efficiently.  The first three summands in \eqref{eq-Contri1} and \eqref{eq-Contri2} are non-negative and on each a quantitative (positive) lower bound will be proved below such that the sum of the three is strictly positive; the last two summands \eqref{eq-Contri3} and \eqref{eq-Contri4} will then be shown to be a perturbation that does not spoil the positivity. Let us start out with a lower bound on 
\begin{align*}
\eqref{eq-Contri2}
\;=\;&
\pi_{\rho'}G_{\rho,\lambda}(D)^2((H^{\ImVar})^*H^{\ImVar}\oplus H^{\ImVar}(H^{\ImVar})^*)G_{\rho,\lambda}(D)^2\pi_{\rho'}^{*}
\\
& \;
-\pi_{\rho'}G_{\rho,\lambda}(D)^2
((H^{\ImVar})^*\oplus H^{\ImVar})[(H^{\ImVar}\oplus (H^{\ImVar})^*),G_{\rho,\lambda}(D)]
G_{\rho,\lambda}(D)\pi_{\rho'}^{*}
\\
& \;-
\pi_{\rho'}G_{\rho,\lambda}(D)
[G_{\rho,\lambda}(D),((H^{\ImVar})^*\oplus H^{\ImVar})]G_{\rho,\lambda}(D)(H^{\ImVar}\oplus (H^{\ImVar})^*) 
G_{\rho,\lambda}(D)\pi_{\rho'}^{*}
\\
\geq \; 
& \; g^{2}\pi_{\rho'}G_{\rho,\lambda}(D)^4\pi_{\rho'}^{*}
\\
& \;
-\pi_{\rho'}G_{\rho,\lambda}(D)^2
((H^{\ImVar})^*\oplus H^{\ImVar})[(H^{\ImVar}\oplus (H^{\ImVar})^*),G_{\rho,\lambda}(D)]
G_{\rho,\lambda}(D)\pi_{\rho'}^{*}
\\
&\; +
\pi_{\rho'}G_{\rho,\lambda}(D)
[(H^{\ImVar})^*\oplus H^{\ImVar}),G_{\rho,\lambda}(D)]G_{\rho,\lambda}(D)(H^{\ImVar}\oplus (H^{\ImVar})^*) 
G_{\rho,\lambda}(D)\pi_{\rho'}^{*}
\\
= \;&\;  g^{2}\pi_{\rho'}G_{\rho,\lambda}(D)^{4}\pi_{\rho'}^{*}
\\
& \;
-\lambda\;\pi_{\rho'}G_{\rho,\lambda}(D)^2
((H^{\ImVar})^*\oplus H^{\ImVar})[H\oplus H^*,G_{\rho}(D)]
G_{\rho,\lambda}(D)\pi_{\rho'}^{*}
\\
&\; +
\lambda\;\pi_{\rho'}G_{\rho,\lambda}(D)
[H\oplus H^*,G_{\rho}(D)]G_{\rho,\lambda}(D)(H^{\ImVar}\oplus (H^{\ImVar})^*) 
G_{\rho,\lambda}(D)\pi_{\rho'}^{*}
\;.
\end{align*}
The first summand is positive and will nicely combine with those in \eqref{eq-Contri1}, the others combine with \eqref{eq-Contri3} and \eqref{eq-Contri4} to an error term
\begin{align*}
E_{\rho,\rho'}(s,\lambda)
\;= \;
& 
\kappa\,\pi_{\rho'}G_{\rho,\lambda}(D)\begin{pmatrix}0 & [H,D_{0}]^{*}\\{}
[H,D_{0}] & 0
\end{pmatrix}G_{\rho,\lambda}(D)\pi_{\rho'}^{*}
\\
& 
\;+\;
\lambda\,
\pi_{\rho'}G_{\rho,\lambda}(D)[H^{*}\oplus H,G_{\rho}(D)]G_{\rho,\lambda}(D)(H^{s}\oplus(H^{s})^{*})G_{\rho,\lambda}(D)\pi_{\rho'}^{*}\\
& 
\;+\;
\lambda\,\pi_{\rho'}G_{\rho,\lambda}(D)^{2}((H^{s})^{*}\oplus H^{s})[G_{\rho}(D),H\oplus H^{*}]G_{\rho,\lambda}(D)\pi_{\rho'}^{*}
\\
& 
\;+\;
2\,s\,\pi_{\rho'}G_{\rho,\lambda}(D)\,\Im m\big[((-H)^{*}\oplus H)(\one-G_{\rho,\lambda}(D)^{2})\big]G_{\rho,\lambda}(D)\pi_{\rho'}^{*}
\;.
\end{align*}
Then, neglecting also the $s^2$-term in \eqref{eq-Contri1},
\begin{align*}
|L_{\kappa,\rho,\rho'}(H;\lambda)-\imath \,s\,1_{\rho'}|^{2}
\;\geq \;& 
\kappa^{2}\,D_{\rho'}^{2}\,+\,g^{2}\pi_{\rho'}G_{\rho,\lambda}(D)^{4}\pi_{\rho'}^{*}
\,+\,
E_{\rho,\rho'}(s,\lambda)
\;.
\end{align*}
Note that this is an inequality for matrices on the finite dimensional space $\Ran(\pi_{\rho'})$. This latter space will be decomposed into $\Ran(\pi_{\frac{\rho}{2}})\oplus\Ran(\pi_{\rho',\frac{\rho}{2}})$ where $\pi_{\rho',\frac{\rho}{2}}=\pi_{\rho'}\ominus\pi_{\frac{\rho}{2}}$. Then the strict positivity of the r.h.s. is proved by providing quantitative positive lower bounds on the two diagonal terms, and then showing the positivity of the $2\times 2$ block matrix is not spoiled by the two off-diagonal terms. Note that the first summands are diagonal in this decomposition, hence the only off-diagonal contribution stems from $E_{\rho,\rho'}(s,\lambda)$.

\vspace{.2cm}

Let us start with the positive term on $\Ran(\pi_{\frac{\rho}{2}})$. As $\pi_{\frac{\rho}{2}}G_{\rho,\lambda}(D)^{4}\pi_{\frac{\rho}{2}}^{*}=\pi_{\frac{\rho}{2}}\pi_{\frac{\rho}{2}}^{*}=\one_{\frac{\rho}{2}}$, one gets
$$
\kappa^{2}\,D_{\frac{\rho}{2}}^{2}\,+\,g^{2}\pi_{\frac{\rho}{2}}G_{\rho,\lambda}(D)^{4}\pi_{\frac{\rho}{2}}^{*}
\;\geq\;
g^2\,\one_{\frac{\rho}{2}}
\;.
$$
The error term $E_{\rho,\rho'}(s,\lambda)$ restricted to $\Ran(\pi_{\frac{\rho}{2}})$ only contains the first three summands because $(\one-G_{\rho,\lambda}(D)^{2})\pi_{\frac{\rho}{2}}^{*}=0$. Thus with $\lambda\leq 1$ and $\|H^\ImVar\|\leq 3\|H\|$ for $|s|\leq 2\|\Im m(H)\|$ and \eqref{eq-TuningBound}, it follows as in \eqref{eq-EBound} that
\begin{align*}
\big\|\pi_{\frac{\rho}{2}}E_{\rho,\rho'}(s,\lambda)\pi_{\frac{\rho}{2}}^{*}\big\|
& 
\;\leq \;
\kappa\|[H,D_{0}]\|+2\lambda\|H^{s}\|\cdot\|[G_{\rho}(D),(H\oplus H^{*})]\|
\;\leq\;
c_\kappa\,{\big(1+\tfrac{48}{c_\rho}\big)}\,g^{2}
\;=\;
\tfrac{3}{4}\,g^2
\;.
\end{align*}
Together one concludes that  $\pi_{\frac{\rho}{2}}|L_{\kappa,\rho,\rho'}(H,\lambda)-\imath s1_{\rho'}|^{2}\pi_{\frac{\rho}{2}}^{*}>\tfrac{1}{4}\,g^2\,\one_{\frac{\rho}{2}}$. Next let us come to the other diagonal part.  Using merely the first term, one has for the positive contribution 
\begin{align*}
\pi_{\rho',\frac{\rho}{2}}
\Big(
\kappa^{2}D_{\rho'}^{2}+g^{2}\pi_{\rho'}G_{\lambda,\rho}(D)^{4}\pi_{\rho'}^{*}
\Big)\pi_{\rho',\frac{\rho}{2}}^{*}
\;\geq\; 
\tfrac{\kappa^{2}\rho^{2}}{4} \,\one_{\rho',\frac{\rho}{2}}
\;\geq\; 
\tfrac{c_\rho^{2}}{4}\,\big(1+\tfrac{\|\Im m(H)\|}{g}\big)^2 \,g^2\,\one_{\rho',\frac{\rho}{2}}
\;.
\end{align*}
Let us next bound the error $E_{\rho,\rho'}(s,\lambda)$ on $\Ran(\pi_{\rho',\frac{\rho}{2}})$. The last summand needs particular care, based on the following identity:
$$
\Im m\big[((-H)^{*}\oplus H)(\one-G_{\rho,\lambda}(D)^{2})\big]
=
(\one -G_{\rho,\lambda}(D)^{2})\Im m((-H)^{*}\oplus H)
+
\frac{1}{2\imath}
[G_{\rho,\lambda}(D)^{2},(-H)^{*}\oplus H]
\;.
$$
Using the fact that $(1-x^{2})x\leq\frac{2}{9}\sqrt{3}\leq \frac{1}{2}$ for all $x\in[0,1]$,
\begin{align*}
\big\|
\pi_{\rho',\frac{\rho}{2}}E_{\rho,\rho'}(s,\lambda)\pi_{\rho',\frac{\rho}{2}}^{*}
\big\|
& 
\;\leq\;
\kappa\|[H,D_{0}]\|+2\lambda\|[(H^{*}\oplus H),G_{\rho}(D)]\|\cdot\|H^{s}\|
\\
&
\qquad
+2|s|\big(\tfrac{1}{2} \|\Im m(H)\|+{\lambda \|[G_{\rho}(D)},(-H)^{*}\oplus H]\|\big)
\\
&
\;\leq \;
\tfrac{3}{4}\,g^2\,+{4\,\|\Im m(H)\| \big(\tfrac{1}{2} \| \Im m(H)\| +\tfrac{8}{\rho}\,N\big)}
\\
&
\;\leq \;
\tfrac{3}{4}\,g^2\,+2\,\|\Im m(H)\|^2+ \tfrac{32\, c_\kappa\,\|\Im m(H)\|\,g^2}{c_\rho\,\| H\|\,(1+ g^{-1} \|\Im m(H)\|)} 
\\
&
\;\leq \;
\tfrac{3}{4}\,g^2\,+2\,\|\Im m(H)\|^2+ {\tfrac{32\,c_\kappa}{c_\rho}\,g^2}
\;,
\end{align*}
where in the the last two steps respectively the bounds in \eqref{eq-HypBound} and $\| \Im m(H)\|\leq \|H\|$ were used. Thus one obtains
$$
\pi_{\rho',\frac{\rho}{2}}|L_{\kappa,\rho,\rho'}(H,\lambda)-\imath\,s\,\one_{\rho'}|^{2}\pi_{\rho',\frac{\rho}{2}}^{*}
\;\geq\;
\big(\tfrac{c_\rho^{2}}{4}\big(1+\tfrac{\|\Im m(H)\|}{g}\big)^2 -\tfrac{3}{4}-2\,\tfrac{\|\Im m(H)\|^2}{g^2}
-{\tfrac{32\,c_\kappa}{c_\rho}}
\big) \,g^2\,
\one_{\rho',\frac{\rho}{2}}
\;.
$$
Finally let us bound the off-diagonal term $\pi_{\frac{\rho}{2}} E_{\rho,\rho'}(s,\lambda)\pi_{\rho',\frac{\rho}{2}}^{*}$. Again by $\pi_{\frac{\rho}{2}}(\one -G_{\rho,\lambda}(D)^{2})=0$, the first summand in the above formula for $\Im m\big[((-H)^{*}\oplus H)(\one-G_{\rho,\lambda}(D)^{2})\big]$ drops out. Hence by the estimate above
$$ 
\big\|
\pi_{\frac{\rho}{2}}E_{\rho,\rho'}(s,\lambda)\pi_{\rho',\frac{\rho}{2}}^{*}
\big\|
\;\leq\;
\tfrac{3}{4}\,g^2 + {\tfrac{32\,c_\kappa}{c_\rho}\,g^2}
\;.
$$
The matrix $\pi_{\rho',\frac{\rho}{2}}E_{\rho,\rho'}(s,\lambda)\pi_{\frac{\rho}{2}}^{*}$ satisfies the same norm bound. 
Therefore in the grading of $\Ran(\pi_{\frac{\rho}{2}})\oplus\Ran(\pi_{\rho',\frac{\rho}{2}})$ one has
$$
|L_{\kappa,\rho,\rho'}(H,\lambda)-\imath\,s\,\one_{\rho'}|^{2}
\;\geq\;
g^2\begin{pmatrix}
\tfrac{1}{4} &   M
\\
M^* & \tfrac{c_\rho^2}{4} (1+\tfrac{\|\Im m(H)\|}{g})^2- \tfrac{3}{4}-
2\,\tfrac{\|\Im m(H)\|^2}{g^2}
-{\tfrac{32\,c_\kappa}{c_\rho}}
\end{pmatrix}
\;,
$$
with off-diagonal error term $M$ satisfying $\|M\|\leq \tfrac{3}{4}+\tfrac{32\,c_\kappa}{c_\rho}$. This is strictly positive as long as 
$$
\tfrac{1}{4}\Big(\tfrac{c_\rho^2}{4}(1+\tfrac{\|\Im m(H)\|}{g})^2 - \tfrac{3}{4}-2\,\tfrac{\|\Im m(H)\|^2}{g^2}
-{\tfrac{32\,c_\kappa}{c_\rho}} \Big)
\;>\;
\Big(\tfrac{3}{4}
+{\tfrac{32\,c_\kappa}{c_\rho}} \Big)^2
\;,
$$
which can readily be verified using {$c_\kappa=\frac{1}{12}$ and $c_\rho=6$}. This concludes the proof of the constancy of the signature.

\section{{Homotopy arguments}}
\label{sec-Proof}

This section proves the equality \eqref{eq-SigInd} which hence concludes the proof of Theorem~\ref{theo-Main}. The strategy will be to homotopically deform the Hamiltonian $H$ and the Riesz projection $P$ into selfadjoint objects for which \eqref{eq-SigInd} is already known by previous works \cite{LS2,DSW}. One then has to show that along those homotopies both sides of the equality \eqref{eq-SigInd} remain constant. Let us start with the index. It is well-known \cite{GVF,DSW} that the Riesz projection $P$ can be deformed into its (selfadjoint) range projection $Q$ by the linear path $t\in[0,1]\mapsto P_{t}= tQ+(1-t)P$ of idempotents. Set $R_t=P_t(\one-(P_t-P_t^*)^2)^{\frac{1}{2}}$. The Fredholm property of $R_tF_0R^*_t+\one-R_tR_t^*$ follows by the argument of the proof of Proposition~\ref{prop-FredProp2} because the commutator $[P_t,F_0]$ is compact (since $[Q,F_0]=[R,F_0]R^*+R[R^*,F_0]$ is compact). Therefore the index is constant along the path so that $\Ind(PF_0P^*|_{\Ran(P)})=\Ind(RF_0R^*+\one-RR^*)=\Ind(QF_0Q+\one-Q)$. Furthermore, by a similar argument one checks that $[D,Q]$ extends to a bounded operator. One can thus use the result of \cite{LS2} (see also \cite{SS1} or Theorem~10.3.1 in \cite{DSW}) applied to the flat-band selfadjoint Hamiltonian $\one-2Q$ to conclude that
$$
\Ind(QF_0Q+\one-Q)
\;=\;
\tfrac{1}{2}\;\Sig(L_{\kappa',\rho'}(\one-2Q))
\;,
$$
where $\kappa'$ can be chosen sufficiently small and $\rho'$ sufficiently large such that bounds similar to \eqref{eq-HypBound} hold. Furthermore, as $[D,Q]$ and $[D,P]$ both extend to bounded operators, so does $[D,P_t]$ for all $t\in[0,1]$. One thus disposes of the bound \eqref{eq-GapSize} for all $t\in [0,1]$, provided that $\kappa'$ and $\rho'$ are chosen sufficiently small and large respectively (note that $g$, $N$ and $\|P\|$ all depend continuously on $t$, and the gap of the spectral localizer remains open). This implies that $\Sig(L_{\kappa',\rho'}(\one-2Q))=\Sig(L_{\kappa',\rho'}(\one-2P))$. Finally, one connects the non-selfadjoint flat band Hamiltonian $\one-2P$ to $H$ by the homotopy $t\in [0,1]\mapsto(1-t)(\one-2P)+tH$, which lies in the set of line-gapped local Hamiltonians. Hence again the line-gap of the spectral localizer remains open along this path and therefore $\Sig(L_{\kappa',\rho'}((1-t)(\one-2P)+tH))$ is constant in $t$. Combining all the above facts, one concludes that
$$
\Ind\big(PF_0P^*|_{\Ran(P)}\big)
\;=\;
\tfrac{1}{2}\;\Sig(L_{\kappa',\rho'}(H))
\;,
$$
for suitable $\kappa'$ and $\rho'$. However, by the results of Section~\ref{sec-SigConst} the signature is constant for all $\kappa>0$ and $\rho>0$ satisfying \eqref{eq-HypBound}. 

\section{Odd-dimensional chiral systems with a line-gap}
\label{sec-OddDim}

Let us briefly explain why the spectral localizer technique for odd-dimensional chiral systems \cite{LS1,SS1,DSW} directly transposes to the study of non-hermitian line-gapped chiral Hamiltonians (local as throughout the paper). Suppose that $H$ and $P$ are given in the spectral representation of $J$:
\begin{equation}
\label{eq-LGH}
J
\;=\;
\begin{pmatrix}
\one & 0 \\ 0 & -\one
\end{pmatrix}
\;,
\qquad
H
\;=\;
\begin{pmatrix}
0 & B \\ A & 0
\end{pmatrix}
\;,
\qquad
P
\;=\;
\frac{1}{2}
\begin{pmatrix}
\one & V^{-1} \\ V & \one
\end{pmatrix}
\;.
\end{equation}
The entries $A$ and $B$ are invertible, and $B=A^*$ for $H$ selfadjoint. The particular form of the entries of $P$ follows from $JPJ=\one-P$, and $V^{-1}=V^*$ is unitary if and only if $P=P^*$. For each of $A$, $B$ and $V$, one computes an odd index pairing, {\it e.g.} $\Ind(EAE+\one-E)$ where $E=\chi(D>0)$ is the Hardy projection. If $H$ and hence $A$ is covariant, then this index is equal to an odd Chern number by an index theorem \cite{PS}.

\begin{proposition}
Let $H$ be a line-gapped chiral Hamiltonian with a Riesz projection $P$ on the spectrum with negative real part. Then there exists a smooth path $t\in [0,1]\mapsto H_t$ of line-gapped and local chiral Hamiltonians such that $H_0=H$ and $H_1=\one-2P$. In particular, the odd index pairings satisfy
$$
\Ind(EAE+\one-E)
\;=\;
-\,
\Ind(EBE+\one-E)
\;=\;
\Ind(EVE+\one-E)
\;.
$$
\end{proposition}

\noindent {\bf Proof.} Let $\gamma$ be a positively oriented path winding once around each point of the spectrum with negative real part so that $P=\oint_\gamma\frac{dz}{2\pi\imath}\,(z\one-H)^{-1}$. The path can be chosen (sufficiently large) such that $-\gamma$ encircles the part of the spectrum with positive real part also with a winding number $1$. Further introduce the interpolating functions $f^\pm_t(z)=(1-t)z\pm t$ which are analytic in the interior of $\gamma$ and $-\gamma$. Hence one can set
$$
H_t
\;=\;
\oint_\gamma\frac{dz}{2\pi\imath}\,f^-_t(z)\,(z\one-H)^{-1}
\;+\;
\oint_{-\gamma}\frac{dz}{2\pi\imath}\,f^+_t(z)\,(z\one-H)^{-1}
\;.
$$
The first summand acts non-trivially merely on the range of $P$, while the second on the range of $\one-P$. The spectral mapping theorem implies that $H_t$ has a line-gap for all $t\in [0,1]$. Furthermore, one readily checks that $JH_tJ=-H_t$. As clearly $H_0=H$ and $H_1=-P+(\one-P)$ the path has all the properties claimed in the statement. This homotopy directly implies that the index pairings of $A$ and $V$ coincide, as do those of $B$ and $V^{-1}$. As those of $V$ and $V^{-1}$ differ by a sign, the claim follows.
\hfill $\Box$

\vspace{.2cm}

The index of $A$ and $B$ can separately be accessed by the selfadjoint odd spectral localizers \cite{LS1,DSW}, but  alternatively one can also use the non-selfadjoint one involving the Hamiltonian:
$$
\Ind(EAE+\one-E)
\;=\;
\frac{1}{2}\;\Sig(L^\od_{\kappa,\rho}(H))
\;,
\qquad
L^\od_{\kappa}(H)
\;=\;
\begin{pmatrix}
\kappa D & B \\ A & -\kappa D
\end{pmatrix}
\;,
$$
provided $\kappa$ and $\rho$ satisfy \eqref{eq-HypBound}.

\appendix

\section{Formulas for the signature}
\label{sec-Signature}

The signature of an $N\times N$ matrix $L$ with no spectrum on the imaginary axis is equal to the difference of the total algebraic multiplicity of all eigenvalues with positive and negative eigenvalues. According to Theorem~\ref{theo-Main}, the signature of the finite volume spectral localizer is the topological invariant of interest. This appendix discusses two ways to access $\Sig(L)$, one via a spectral flow and one via a winding number. Let us begin by recalling ({\it e.g.} Section~1.6  of \cite{DSW}) that for a continuous path $t\in [0,1]\mapsto L_t$ of matrices such that the endpoints $L_0$ and $L_1$ have no spectrum on the imaginary axis, the spectral flow of the path is given by
\begin{equation}
\label{eq-SFSIG}
\SF(t\in [0,1]\mapsto L_t)
\;=\;
\frac{1}{2}\big(\Sig(L_1)\,-\,\Sig(L_0)\big)
\;.
\end{equation}
Let us stress  that this is in general {\it not} the spectral flow of the path $t\in [0,1]\mapsto \Re e(L_t)=\frac{1}{2}(L_t+L_t^*)$. The formula \eqref{eq-SFSIG} can be used to compute the signature $\Sig(L)$ if one chooses a suitable path with $L_1=L$ and for which the signature $\Sig(L_0)$ is known. An example of such a path is certainly given by $L_t=L+2(1-t)\|L\|$, for which $\Sig(L_0)=N$. In Section~\ref{sec-Numerics} rather exhibits a path for which $\Sig(L_0)=0$. Such paths are advantageous (in numerical applications) if the signature of $L$ is small compared to the size $N$.  The spectral flow of the path can be obtained numerically by computing the low-lying spectrum of $L_t$ for all $t\in [0,1]$ (of course, the path is discretized and typical paths are actually analytic in $t$). 

\vspace{.2cm}

Another formula for the signature is known as the Routh-Hurewitz theorem. As shows the short proof below, it is a basic consequence of the argument principle. It is a way to access the non-hermitian signature as a suitable winding number. Again this is potentially of use for numerics in situations where the signature is small compared to the size of the matrix so that the winding number appearing below is also small. While this formula is not implemented in the present work, it is certainly of theoretical interest in this context.

\begin{proposition}
\label{prop-SigRot}
Let $L$ be an $N\times N$ matrix with a line-gap on the imaginary axis. Then its half-signature is given by
\begin{align}
\label{eq-SigRot}
\frac{1}{2}\;\Sig(L)
&
\;=\;
\int^{\infty}_{-\infty} \frac{ds}{2\pi\imath}\;\partial_s\,
\ln\big(\det(L+\imath \,s\,\one)\big)
\\
&
\;=\;
\label{eq-SigRot3}
\int^{\infty}_{-\infty} \frac{ds}{2\pi}\;\frac{1}{1+s^2}\;\Tr\big((\one+\imath \,s\, L)(L+\imath \,s\,\one)^{-1}\big)
\;.
\end{align}
\end{proposition}

\noindent {\bf Proof.} The characteristic polynomial $z\in\CM\mapsto \det(L-z\one)$ is analytic and of the form $\det(L-z\one)=(-z)^N+\Oo(|z|^{N-1})$. Let us introduce the meromorphic function
$$
f(z)
\;=\;
\frac{1}{2\pi\imath}\;\frac{\partial_z\,\det(L-z\one)}{\det(L-z\one)}
\;.
$$
Even though not used in the following, let us note that the fundamental theorem of algebra and the argument principle implies that
$$
N
\;=\;
\oint_{\Gamma_R} dz\,f(z)
\;,
$$
where $\Gamma_R$ is a positively oriented circle of sufficiently large radius $R$, centered at the origin. Let us split $\Gamma_R=\Gamma_R^+\,+\,\Gamma_R^-$ into the half-circle with positive and negative real part. Then an explicit computation shows
$$
\frac{N}{2}
\;=\;
\lim_{R\to\infty}
\oint_{\Gamma^\pm_R}
dz\,f(z)
\;.
$$
Furthermore, let $\Gamma_R^0$ be the path $s\in[-R,R]\mapsto \imath s\in\CM$. By hypothesis $f$ has no pole on $\Gamma_R^0$. If now $N_+$ and $N_-$ denote the number of zeros of $\det(L-z\one)$ (counted with their multiplicity) on the right and left half-plane respectively, then by the argument principle
$$
N_- 
\;=\;
\oint_{\Gamma^-_R\,+\,\Gamma_R^0}
dz\,f(z)
\;,
\qquad
N_+
\;=\;
\oint_{\Gamma^+_R\,-\,\Gamma_R^0}
dz\,f(z)
\;.
$$
Taking the difference then shows
$$
N_+\,-\,N_-
\;=\;
\oint_{\Gamma^+_R}
dz\,f(z)
\;-\;
\oint_{\Gamma^-_R}
dz\,f(z)
\;-\;2
\oint_{\Gamma_R^0}
dz\,f(z)
\;.
$$
Taking the limit $R\to\infty$ now implies the first equality \eqref{eq-SigRot} (note that the sign is obtained by the change of orientation in the statement). Using the identity $\ln\det=\Tr \ln$ and then deriving directly implies 
$$
\frac{1}{2}\;\Sig(L)
\;=\;
\lim_{R\to\infty} \int^R_{-R}\frac{ds}{2\pi}\;\Tr\big((L+\imath \,s\,\one)^{-1}\big)
\;.
$$
Now the integrand only decays as $\frac{1}{s}$ at $s\to\pm\infty$ and hence is not integrable. However, one can regularize with
$$
\int^R_{-R}\frac{ds}{2\pi}\;\frac{\imath s}{1+s^2}
\;=\;0
\;.
$$
Due to
$$
\Tr\Big((L+\imath \,s\,\one)^{-1}+\frac{\imath s}{1+s^2}\,\one\Big)
\;=\;
\frac{1}{1+s^2}\;\Tr\big((\one+\imath \,s\,L)(L+\imath \,s\,\one)^{-1}\big)
\;,
$$
this leads to the second equality \eqref{eq-SigRot3} because the integral is now absolutely convergent.
\hfill $\Box$

\vspace{.2cm}

\noindent {\bf Acknowledgements:} We thank Enrique Zuazua for reminding us of the Routh criterion. This work was supported by the DFG grant SCHU 1358/6-2.  A.C.\ acknowledges support from the Center for Integrated Nanotechnologies, an Office of Science User Facility operated for the U.S.\ Department of Energy (DOE) Office of Science, and the Laboratory Directed Research and Development program at Sandia National Laboratories. Sandia National Laboratories is a multimission laboratory managed and operated by National Technology \& Engineering Solutions of Sandia, LLC, a wholly owned subsidiary of Honeywell International, Inc., for the U.S.\ DOE's National Nuclear Security Administration under contract DE-NA-0003525. The views expressed in the article do not necessarily represent the views of the U.S.\ DOE or the United States Government.


\end{document}